# AN APPROACH TO ADAPTIVE MICROLEARNING IN HIGHER EDUCATION

Ovidiu Gherman[1], Cristina Elena Turcu[1], Corneliu Octavian Turcu[1]

[1]*Stefan cel Mare University of Suceava, Romania*

## Abstract

Current changes in society and the education system, cumulated with the accelerated development of new technologies, entail inherent changes in the educational process. Numerous studies have shown that the pandemic has forced a rapid transformation of higher education. Thus, if before the pandemic digital technologies were used to supplement the learning process, now they are the main means of learning delivery. In addition, as previous research has shown, new pedagogical strategies and new ways of teaching and learning are needed for the current generation of students, the so-called Generation Z, to acquire new knowledge and develop new skills. In this necessary evolution of the educational process, a possible solution to increase the effectiveness of the learning process for the Generation Z students is to use microlearning to extend the traditional ways of learning. Many studies have shown that microlearning, based on how today's students learn and memorize, facilitates learning. In recent years there has been a growing trend in ther use of microlearning in the educational process. But, in order to be effective, this approach must allow the individual knowledge building, by indicating a guiding direction of the optimal path to achieve the proposed objectives. The benefits resulting from introducing adaptive microlearning in higher education are discussed. We propose a system for personalized learning using microlearning, which provides support and guidance to students based on their individual needs, in order to increase their interest in learning, but also to compensate for various deficiencies in their educational background. We also present a case study from the higher education sector, taking into account the discipline of Computer Graphics, provided in the curriculum of the Computers study program at our University. Feedback from students and data collected during the semester as a result of the students' behavioural analysis and their real learning motivations will be used to improve the proposed system. Further pedagogical research should also be conducted to demonstrate the improvement of students' learning experiences.

## 1 INTRODUCTION

Higher education must face the challenges brought by the technological developments, but also by societal changes. Thus, with admission into higher education of Generation Z students, other requirements must be met, as their learning expectations are different from those of previous generations [1]. Many studies highlight that the appropriate type of educational support for Generation Z students may be different from the support needed by previous generations. Thus, Gen Z wants learning experiences adapted to them and easy to consume - from any device, whenever and wherever they want. Some of the higher education changes have already taken place, especially in the context of the pandemic, as Global Higher Education Research presented in [2]. Thus, in response to the pandemic, many higher education institutions have already implemented or intend to implement more flexible learning options for when and how students can learn. The question is what teaching methods and tools could be used to restructure a course so that the effects will be noticeable.

In this necessary evolution of the educational process, a possible approach to increase the efficiency of the learning process applicable to Gen-Z is to use microlearning as a supplement to traditional teaching methods. Numerous studies have shown that microlearning – based on today's students learning and memorisation patterns- allows for a better educational outcome. In the last years there is an increasing trend towards using microlearning in education. But – to be efficient – this solution must allow for individual increase in knowledge while at the same time tracing the optimal path towards the educational objectives.

Generation Z is focused more on quickness than accuracy. In order to increase students' efficiency, a solution is to consider microlearning adoption as they can use it to learn anywhere and anytime on any device.

The authors of this paper have already successfully adopted microlearning in their teaching activities. In [3] they presented a case study of integrating microlearning units in the Elements of Computer Graphics (ECG) course, provided in the first semester of the 3rd year in the curriculum of the Computers undergraduate study program, in the field of Computers and Information Technology at the Faculty of Electrical Engineering and Computer Science from Ștefan cel Mare University of Suceava.

This paper presents a solution for adaptive microlearning with the aim of improving individual educational experiences of students. The authors propose a system for personalised learning using microlearning, system-oriented towards student guidance using individual learning necessities, increasing their interest in learning but also to compensate for deficiencies in their previous education.

The paper starts by presenting the research methodology particularly used by the authors of this paper. Then a comprehensive introduction to microlearning and adaptive microlearning is provided. In the next sections, we put adaptive microlearning in the context of learning and teaching and then focus on a specific application in a computer graphics course. Following this, we present future research directions and conclusions of this paper.

## 2 METHODOLOGY

In this paper, the authors try to respond to a couple of research questions centered around the microlearning concept:

    Q1: *What is microlearning?*

    Q2: *What are the applications of microlearning in higher education?*

    Q3: *How can microlearning be applied effectively in personalized learning?*

This relatively new branch of e-learning is making inroads in higher education (at least in the last few years), so it is important to evaluate its suitability for specific situations.

## 3 CONCEPTUAL FOUNDATIONS OF MICROLEARNING

The term "microlearning" generally describes "*all about getting your eLearning in small doses, as tiny bursts of training material that you can comprehend in a short time (contrast with the hefty tomes you had to read at school to study a subject or the typical content-heavy eLearning class — which would be classified as "macro" learning)*" [4].

According to [5], [6], small activities and short-term goals could be considered to schedule very effective learnings. These micro-content-based activities are then used in wider knowledge or in long-term learning. According to [7] there are many concepts and versions of microlearning. These versions have some common features, including microcontent and short learning time (i.e., no longer than 15 minutes) [8].

An examination of the scientific literature demonstrates that microlearning was applied in various higher education specialisations, such as, engineering [9], textiles [10], tourism [11], health [12]-[15], educational sciences [16], etc. In the beginning, microlearning was used especially in private enterprises while being usually avoided in academia. But in the last decade various research was published, research that highlighted the advantages of the microlearning in the educational process, both in the private sector and in the academia (where the adoption of this technique is increasing). According to the findings of various studies regarding the publication trend of microlearning, "the microlearning could mature and develop into a critical mainstream issue in the future or become a major trend in its own right. As a result, researchers in the field should consider microlearning as a promising research direction" [17]. Online learning could play a very important role given the uncertainties surrounding the coronavirus pandemic. Microlearning could be one of the responses to the increased demand for online learning opportunities.

On the other hand, the design and development of microlearning units can put pressure on the teacher. Given the fast pace of knowledge acquisition in many areas, the Internet can be a viable solution that can offer free study resources, many valuable, that can be successfully used in a microlearning context (as open education resources - OER). Unfortunately, these microlearning units are usually un-organised. In other cases, they are sub-par (as content) or presented in a manner that make them unfit for presentation from a pedagogical aspect [18]. The right approach to select, organise and – if necessary – to remake the units is fundamental in this approach to make higher education more effective [19].

## 4    CASE STUDY

The recommendation system is proposed to be applied in the Elements of Computer Graphics (ECG) course. The course considered herein is a required course for students in the Computer program at our university. This course is offered over one 14-week semester. Due to pandemic, the course transitioned to full online instruction in the first semester of 2020-2021 year. The first seven laboratories were also held online and, in the last seven weeks students must attend in-class laboratories.

This course has a series of academic prerequisites derived from other courses studied in the faculty's curricula. The main purpose of our approach is to identify weak points in the student's understanding of various concepts in computer graphics field and to correct it via targeted, dedicated "packages" of informational content. In this way, we surmise that the student will have a better understanding of the subject, and the learning process will be more effective.

In a microlearning environment, it is important to identify gaps in knowledge of the student, caused usually by either the lack of understanding of fundamental concepts that are important for the current course or the lack of certain concepts which, perhaps, have not been presented to the student in the first place (due to insufficient education or other factors in the past). Every advanced topic has certain prerequisites that must be satisfied to allow building new knowledge and operational knowledge on top of a solid foundation. Given that certain courses may be prerequisites for further topics, a solid understanding of the basics is always recommended.

As a first step, the identification of possible deficiencies in understanding the topic at hand is a very important component of this process. In this regard we propose a solution in order to:

    a) identify the deficiencies in the prerequisites for the Elements of Computer Graphics course,

    b) provide the student with relevant information regarding the identified problem,

    c) ensure that the student understood and assimilated the required information, allowing him to continue building new knowledge.

This process is repeated periodically, so that all potential knowledge gaps are identified and solved. This process may also be effective, due to the fact that is based around the student; with each student being unique and having certain variation, non-overlapping, in the understanding of the prerequisites of the topic, the educational process can be better targeted for each individual resulting in a positive educational outcome. Moreover, if the microlearning units can be offered automatically based on a reference engine that will generate a proper feedback for the student's responses (regarding the lack of knowledge and the remedial units that can be studied to fill the gaps), the educational effect could be augmented because the student can retain his or her privacy and follow the remedial units on its own speed and isolated from any emotional pressure from peers or teachers (this being especially useful for introverts).

We present a conceptual system design for the proposed solution. Fig.1 presents the overall workflow of the proposed system.

As we observe, the process must start with an initial evaluation, to test the general knowledge level in the identified prerequisites for the current topic. With the help of an online quiz, students can self-assess themselves with the purpose of evaluating their level of readiness regarding mathematical background required for this course. Their results are saved in the Learning Management System (LMS). In order to recommend the microlearning units, the proposed system is based on the analysis of this information saved in the LMS.

We propose that once the student completes the evaluation, based on the grade obtained and on the wrong responses registered by the system, he or she can be classified in 3 distinct categories: *pass*, *pass with remediation* and *fail*. If the grade is below a certain given (minimal) threshold (the evaluation categories can be weighted in grading as to highlight their importance), it can be considered that the student failed and fundamental prerequisites were lacking; this will trigger the next stage in the process, where appropriate microlearning modules will be offered (in various forms suitable to the topic at hand – as training video, premade projects, presentations, etc.) to the student. If the grade is above a maximal threshold, the student does not need remedial action and can continue with its regular course material. If the grade is somewhere between the minimal and the maximal threshold, this is a sign that some remediation work is needed, but the student's grasp of prerequisites is quite solid. In this situation, recommendations may be given as to what improvements might be made, based on his/her responses, without any follow-up evaluation being necessary; the individual will study the unit if and when will consider necessary. Students have access to any microlearning unit by themselves, any time and from anywhere.

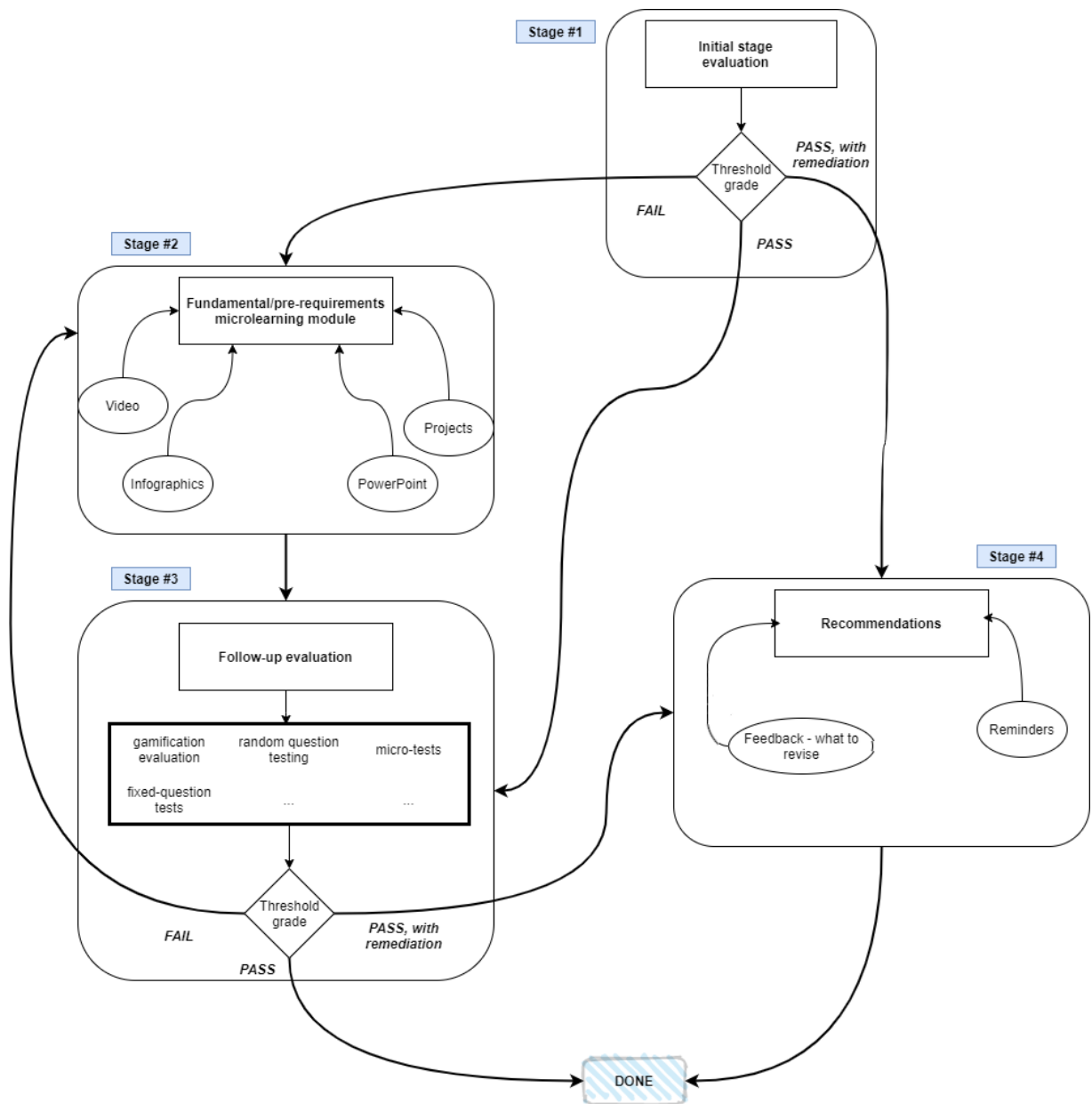

*Figure 1. The workflow of the proposed solution.*

In the second stage, if the student failed the test, a proper microlearning unit must be selected and presented; the student will have to pass a follow-up evaluation, and – if he/she fails – he/she will have to retake the unit until the knowledge is retained. The follow-up evaluation can take many forms, suitable to the content studied in the course and its presentation manner: gamification, random questions testing, multiple answers test, micro-tests, etc. This method is oriented towards Science, Technology, Engineering, and Mathematics (STEM) students. On each iteration, the follow-up evaluation will be used to classify the student in one of the 3 categories: fail (requiring a retake of microlearning unit), pass (meaning that the student can continue with the regular content in the course) and pass with remediation. The last case (as stated previously) will make use of the recommendation engine to offer feedback oriented towards the zones where the student failed to respond correctly. This feedback will highlight what the student must revise and will set periodic reminders for the individual until the goal is satisfied.

At the foundational level of our system sits the feedback engine. This engine provides recommendations which the student can use to revise more efficiently and more effectively. The recommendation system must make the connection between the fundamental requirements and the gaps in the student's knowledge. A useful approach that can be used is the concept of mind map. The research shows that the students are helped by this approach [20], [21] – especially in the technical fields where the knowledge volume is extremely high. There are various tools (oriented mostly towards e-learning) that use mind maps to structure the content for easy parsing and identification and to establish a chain of knowledge in order to deepen the knowledge of the students [22]. Experimental results suggest that this approach is effective in the e-learning domain [23].

Mind maps allow us to present certain related concepts, in a manner easy to understand both by humans and machines. In Fig. 2 we present an example of the mind map representing a fragment of the fundamental knowledge in the topic of Elements of Computer Graphics, and the prerequisites that must be satisfied for a solid understanding of that fragment.

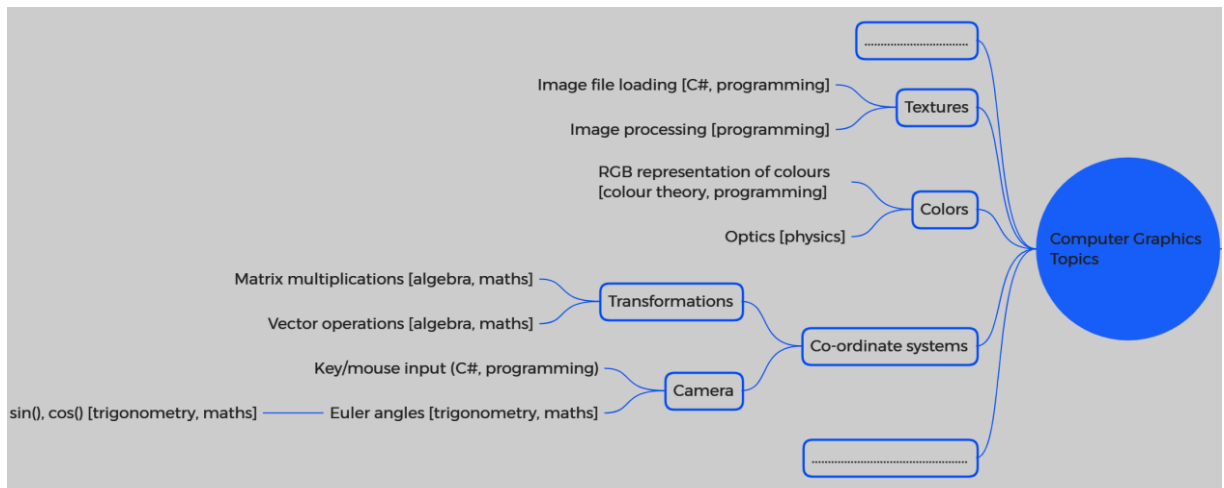

*Figure 2. Mind map representation of prerequisites (fragment).*

This kind of approach is very flexible because it can describe the process to humans (the diagram – being drawn – can be easily understood by students, highlighting the relations between various microlearning units) but also can be machine-readable. Once the mind map is written, the output can be parsed as a graph, then analysed internally from that perspective – including programmatically.

The recommendation engine will use – for example – the mind map to associate the microlearning units with the correct category, identifying the gaps in student's knowledge. Also, the recommendations can be made after the same principle, via association driven by the mind map. The algorithm that can drive this recommendation engine can be described as follows:

```
start
      get(list_of_wrong_answers);
      foreach(element in list_of_wrong_answers)
            assoc = get.category(element);
            list_of_microunits = search(assoc, mindmap);

            recommend(list_of_microunits);
            set(reminder);
stop
```

If a student's grade falls within a certain predefined range, the recommendation engine will start. Once the recommendation is set, the operation will be registered internally (it will log the student's ID, the date of the event and the recommended topics for further analysis). The identity of the student will be kept anonymous (as not to influence his/her participation in the program) but the educational progress of the person behind the ID can be monitored for signs of improvement. In fact, our solution becomes a personalized learning system, allowing for providing microlearning units for students according to the identified learning path of each of them.

This type of logging is important for another reason: evaluating the quality of the content. Since the most-recommended units are in "high demand", special attention must be offered to the quality and segmentation of information in those units. Student could offer feedback for the usefulness and quality of the content (the method to extract the feedback must be simple and effective to use so an in-depth questionnaire is not recommended) so that the system can highlight the section most needed to rework/restructure. Such analytic data can be used to further refine the system.

## 4.1 Results

The main concern of teaching students is how well they will learn learn the material taught. In this regard, we tried to evaluate the impact of using the proposed system on student learning. All tests designed to support learning are available online, therefore each student can self-assess himself through the website. Moreover, the knowledge related to a microlearning unit is validated through a practical assessment.

Feedback received from students has shown that most of the students enrolled in this course were satisfied with the learning experience provided through microlearning.

In order to deeply analyse benefits and problems of using the proposed system, we took into account two methods for assessing students. Thus, the new knowledge acquired has been assessed through multiple-choice questions that are graded automatically. Data, obtained from two different cohorts of third-year students who took this computer graphics course in two consecutive years, illustrates the significant improvement in student understanding.

We asked students to demonstrate understanding as individuals through the ability to apply the newly assimilated knowledge in the development of software applications in order to successfully complete their homework. These applications have been evaluated by the teachers according to the same grading criteria as the previous year. The student grades were compared with those obtained by students enrolled in the previous year. Data collected over three consecutive years showed a statistically significant improvement in learning outcomes.

However, it should be noted that by the time this paper was written, a partial assessment of learning had been made. Thus, the analysis of the students' results did not include the final grades obtained in the evaluation of the entire activity during the first semester of this academic year. Also, the grades of the final exams could not be considered.

But the partial results confirm that the proposed system is a good choice for microlearning management and encourage us to continue with further research.

## 5 FUTURE RESEARCH

For future research, we propose to investigate the following important question: *How can we extend the use of the proposed system in the future?*

With the intent to foster learning activities in computer graphics education that encourage active student participation, we propose to continue developing other microlearning units and to include various interactions with the students, in order to improve its effectiveness. The developed microlearning units and the proposed system could also be used in the lifelong learning process of adults, for example the so-called Generation Y or millennials.

Emerging technologies in the field of artificial intelligence can provide support for the development of the proposed system, considering different educational scenarios for students. Student test results could be used for data analysis to allow predictions about students who could become at risk, thus enabling timely and personalized interventions. The authors will also consider the development and integration of a smart chatbot or conversational assistant to ensure interaction with students in natural language.

Finally, this approach could extend to other STEM topics, allowing a better educational outcome across multiple technical educational areas. If a platform is set up in such a way that is easy to use by the teacher and useful to the students, it could encourage the adoption of microlearning techniques and methods as teaching tools.

Looking forward, a growing learning paradigm, nanolearning, is gaining popularity as a new branch derived from e-learning.

## 6    CONCLUSIONS

In recent years there has been an important evolution of higher education, by adapting to new generations of students and labour market needs, but also by adopting emerging technologies. The pandemic revealed the gap between digital technologies and digital education, but at the same time gave a boost to digital education. The challenge is to integrate new technologies into current learning contexts and use them to provide high quality education for all. Consequently, teachers can no longer rely on traditional ways of study. Generation Z has a focus more on quickness than accuracy. In order to increase students' efficiency, a solution is to consider microlearning adoption as they can use it to learn anywhere and anytime and on any device. Various publications estimate that in the future microlearning could play an increasing role in the educational process. Although considerable progress has been made, it is still difficult to effectively use microlearning units in higher education.

In this paper, we report an experience of using microlearning in a computer graphics course. The purpose of the microlearning units was to get students up to speed as quickly as possible in the field of computer graphics. In this regard, we provided students with a map of microlearning units for this course, which was designed to be as convenient as possible, but without sacrificing any of the essentials. Furthermore, we have implemented a solution whereby each student receives specific advice on the microlearning units to be covered adapted to his or her own learning path. But it is important that students actively participate in the proposed activities in order to develop their key abilities for computer graphics.

The positive effects of adaptive microlearning usage could encourage other teachers to rethink their traditional course structures in order to increase students' motivation.

## ACKNOWLEDGEMENTS

This research was supported by the project "Integrated center for research, development and innovation in Advanced Materials, Nanotechnologies, and Distributed Systems for fabrication and control," No. 671/09.04.2015, Sectoral Operational Program for Increase of the Economic Competitiveness, co-funded by the European Regional Development Fund.

## REFERENCES

[1]     *, Generation Z, Center for Innovative Teaching and Learning, Instructional Guide, Northern Illinois University, 2020, Available at: https://www.niu.edu/citl/resources/guides/instructional-guide/generation-z.shtml

[2]     *, The Evolution of Higher Education: 5 Global Trends to Watch, Global Higher Education Research Snapshot, Salesforce.org, November 2020, Available at: https://www.visualcapitalist.com/the-evolution-of-higher-education-5-global-trends-to-watch/, https://www.salesforce.org/higher-education-research-trends/?d=&nc=

[3]     C.E. Turcu, C.O. Turcu, O. Gherman, Microlearning as a Facilitator of Learning Delivery, The 15th International Conference on Virtual Learning ICVL 2020, October 2020, University of Bucharest, Romania, ISSN: 1844-8933, Available at: http://www.c3.icvl.eu/2020/papers2020.

[4]     Andriotis, N., Everything you wanted to know about microlearning (but were afraid to ask), 2016, Available at: https://www.efrontlearning.com.

[5]     Hug, Theo, and Norm Friesen. "Outline of a Microlearning Agenda." Didactics of Microlearning. Concepts, Discourses and Examples, 15–31, 2007.

[6] Hug T. Micro Learning and Narration. Exploring possibilities of utilization of narrations and storytelling for the designing of" micro units" and didactical micro-learning arrangements. In: fourth Media in Transition conference. 2005.

[7] Hug, T., Micro Learning and Narration. In Fourth Media in Transition Conference, May (pp. 6-8). Cambridge: MIT Press. 2005.

[8] Buchem, I. and Hamelmann, H., "Microlearning: a strategy for ongoing professional development", eLearning Papers, Vol. 21 No. 7, pp. 1-15, 2010.

[9] Said, I. and Çavuş, M., "ALU design by VHDL using FPGA technology and micro learning in engineering education", British Journal of Computer, Networking and Information Technology, Vol. 1 No. 1, pp. 1-18, 2018.

[10] Buhu, A. and Buhu, L., The Applications of Microlearning in Higher Education in Textiles, eLearning and Software for Education, Vol. 3, pp. 373-376, 2019.

[11] Dolasinski, M. J., & Reynolds, J., Microlearning: A New Learning Model. Journal of Hospitality & Tourism Research, 44(3), 551-561, 2020.

[12] Wang, C., Bakhet, M., Roberts, D., Gnani, S., & El-Osta, A., The efficacy of microlearning in improving self-care capability: a systematic review of the literature. Public Health, 186, 286-296, 2020.

[13] Gawlik, K., Guo, J., Tan, A., & Overcash, J., Incorporating a Microlearning Wellness Intervention Into Nursing Student Curricula. Nurse Educator, 46(1), 49-53, 2020.

[14] De Gagne, J.C., Park, H.K., Hall, K., Woodward, A., Yamane, S. and Kim, S.S., "Microlearning in health professions education: scoping review", JMIR Medical Education, Vol. 5 No. 2, 2019.

[15] De Gagne, J.C., Woodward, A., Park, H.K., Sun, H. and Yamane, S.S., "Microlearning in health professions education: a scoping review protocol", JBI Database of Systematic Reviews and Implementation Reports, Vol. 17 No. 6, pp. 1018-1025, 2019.

[16] Allela, M. A., Ogange, B. O., Junaid, M. I., & Charles, P. B., Effectiveness of Multimodal Microlearning for In-service Teacher Training. Journal of Learning for Development, 7(3), 384-398, 2020.

[17] Leong K, Sung A, Au D, Blanchard C. A review of the trend of microlearning. Journal of Work-Applied Management. 2020.

[18] Sun, Geng & Cui, Tingru & Beydoun, Ghassan & Chen, Shiping & Dong, Fang & Xu, Dongming & Shen, Jun., Towards Massive Data and Sparse Data in Adaptive Micro Open Educational Resource Recommendation: A Study on Semantic Knowledge Base Construction and Cold Start Problem. Sustainability. 9. 898. 10.3390/su9060898, 2017.

[19] J. Lin, G. Sun, J. Shen, T. Cui, P. Yu, D. Xu, et al. "A Survey of Segmentation Annotation and Recommendation Techniques in Micro Learning for Next Generation of OER", 2019 IEEE 23rd International Conference on Computer Supported Cooperative Work in Design (CSCWD), pp. 152-157, 2019.

[20] Yu, Wen-jing & Wang, Qi & Jia, Guang-tao & Gao, Zhi-qin & Zhao, Chun-ling., Application of Mindmap in the Teaching of Cell Biology of Medical Students. DEStech Transactions on Social Science, Education and Human Science, 10.12783/dtssehs/meit2017/12824, 2017.

[21] Lee, H.J. & Messom, C.H., Adapting engineering education for a mindmap-based digital textbook: To reduce the skill gap between what the industry demands and what academia delivers. 916-924, 2011.

[22] Gao Na and Li Liang, "The Curriculum Integration Thought Based on the Mindmap", Proceedings of the Fifth International Symposium – Education Management and Knowledge Innovation Engineering, pp. 274-277, 5th International Symposium on Education Management and Knowledge Innovation Engineering, Melbourne, AUSTRALIA, APR 20-22, 2012, ISBN 978-0-646-57923-8, 2012.

[23] Jiang, Z. Y., The Empirical Research of Network Teaching System Based on Webquest and Mindmap. Applied Mechanics and Materials, 475–476, 1235–12392013. https://doi.org/10.4028/www.scientific.net/amm.475-476.1235, 2013.